\documentclass[useAMS]{mn2e}
\usepackage{times}
\usepackage{rotating}
\long\def\symbolfootnote[#1]#2{\begingroup%
\def\thefootnote{\fnsymbol{footnote}}\footnote[#1]{#2}\endgroup}
\newcommand{\gae}{\lower 2pt \hbox{$\, \buildrel {\scriptstyle >}\over {\scriptstyle \sim}\,$}}
\newcommand{\lae}{\lower 2pt \hbox{$\, \buildrel {\scriptstyle <}\over {\scriptstyle \sim}\,$}}
\newcommand{\aprop}{\lower 2pt \hbox{$\, \buildrel {\scriptstyle \propto}\over {\scriptstyle \sim}\,$}}

\begin{document}

\title[Delayed arrival of high energy photons from GRBs]
{Magnetic jet model for GRBs and the delayed arrival of $>$100 MeV photons}

\author[Bo\v snjak \& Kumar]{\v Z. Bo\v snjak$^1$\thanks
{E-mail: zeljka.bosnjak@cea.fr,~~~ pk@astro.as.utexas.edu}
and P. Kumar$^{2}$\footnotemark[1] \\
$^{1}$AIM (UMR 7158 CEA/DSM-CNRS-Universit\' e Paris Diderot) Irfu/Service
d'Astrophysique,  Saclay, 91191 Gif-sur-Yvette Cedex,  France\\
$^{2}$Department of Astronomy, University of Texas at Austin, Austin,
TX 78712, USA}

\date{Accepted ; Received ; in original form 2011 July 20}

\maketitle

\begin{abstract}
\noindent
 Photons of energy larger than 100 MeV from long-GRBs arrive a few seconds
after $<10$ MeV photons do. We show that this delay is a natural consequence 
of a magnetic dominated relativistic jet.
The much slower acceleration of a magnetic jet with radius (compared 
with a hot baryonic outflow) results in high energy gamma-ray photons to
be converted to electron-positron pairs out to a larger radius 
whereas lower energy gamma-rays of energy less than $\sim 10$MeV can 
escape when the jet crosses the Thomson-photosphere. The resulting delay
for the arrival of high energy photons is found to be 
similar to the value observed by the Fermi satellite for a number of
GRBs. A prediction of this model is that the delay should increase with
photon energy (E) as $\sim E^{0.17}$ for $E>$100 MeV. 
The delay depends almost linearly on burst redshift, and on the distance
from the central compact object where the jet is launched ($R_0$). Therefore,
the delay in arrival of $>$10$^2$MeV photons can be used to estimate 
burst redshift if the magnetic jet model for gamma-ray generation is 
correct and $R_0$ is roughly the same for long-GRBs.

\end{abstract}


\section{Introduction}

The Fermi satellite has detected 17 GRBs with photons of energy $>$100MeV
in the first ~3 years of operation. The high energy $\gamma$-ray radiation
for most of these bursts detected by the LAT (Large Area Telescope)
 instrument aboard the Fermi satellite
shows two interesting features (Omodei et al. 2009):  (1) The first
$>$100 MeV photon arrives later than the first lower energy photon
($\lae$1 MeV) detected by the GBM (Gamma-ray Burst Monitor) -- Abdo et 
al. 2009a,b, 2010, Ackermann et al. 2010 \& 2011  (2) $>$100
 MeV radiation lasts for a much longer time compared to the burst duration
in the sub-MeV band (Abdo et al. 2009a, 2010).

It is natural to expect radiation lasting for a time duration longer
than the prompt GRB burst duration when it is produced in
the external shock; external shock results when relativistic ejecta from
a GRB runs into the surrounding medium. In fact the Fermi/LAT data 
($>$10$^2$MeV) --- after the prompt GRB phase, i.e. $t\gae30$s --- is 
found to be consistent with the expectation of the external shock model both 
in regards to the absolute value of the flux at 100 MeV and its temporal
evolution (Kumar \& Barniol Duran, 2009, 2010). Moreover, the external shock 
model provides an excellent fit for the entire observed data --- high-energy 
$\gamma$-rays, x-ray, optical and radio frequencies --- in the time interval
of $\sim30$s and a week (Kumar \& Barniol Duran, 2010). These agreements
between the data and the theoretical model provide a compelling case 
that the observed Fermi/LAT photons for $t\gae30$s originate in the 
external shock. However, the external shock model encounters a few 
problems explaining the Fermi/LT data during the prompt phase ($t\lae30$s).

The LAT data appears to show short time scale variability --- although 
it is unclear if this is statistically significant considering the 
relatively small number of photons detected at high energies --- which 
seems to be correlated with the sub-MeV lightcurve. If the lower energy 
$\gamma$-rays detected by Fermi/GBM are produced by a mechanism distinct from 
the external shock --- as suggested by many lines of evidences
e.g. Piran (2004), Zhang (2007) --- then the implication of this correlation 
(if real) is that the $>$10$^2$MeV photons generated during the prompt phase 
might also be produced by the same mechanism.

Another intriguing feature of the data obtained by Fermi is that the
spectrum in the energy interval 8 keV -- 10 GeV, during the prompt phase, 
can be fitted with a single Band function for most GRBs\footnote{For three
bursts an additional power law component extending from $\sim10$keV
to 10 GeV energies is required.}. This suggests that photons in the 
entire Fermi energy band are likely produced by a single mechanism.

A number of mechanisms have been proposed for the generation of 
high energy photons observed from GRBs such as the 
photo-pion, proton synchrotron, inverse-Compton, SSC etc. 
(see Gupta \& Zhang, 2007, Fan \& Piran, 2008 for recent reviews). 
A number of proposals have been put forward to explain the
observed delay between the Fermi/LAT and GBM lightcurves, eg. 
external shock (Kumar \& Barniol Duran, 2009; Ghisellini et al. 2010;
Ghirlanda et al. 2010; Barniol Duran \& Kumar, 2011), 
proton synchrotron radiation \& photo-pion process (Razzaque et al. 2010,
Asano et al. 2009), IC scattering of photospheric or cocoon thermal radiation
by electrons accelerated in internal shocks (Toma et al. 2009 \& 2010)
  neutron-proton collisions (Beloborodov, 2010; Vurm et al. 2011; M\'esz\'aros 
and Rees, 2011), IC scatterings in internal shocks (Bosnjak et al, 2011).

Whenever higher energy photons are produced at larger radii than 
lower energy $\gamma$-rays we expect a delay for the arrival of high 
 energy photons. Another possible way that a delay could
arise is if the radiation is produced while the jet is undergoing
acceleration; in this case high energy photons are trapped -- converted
to $e^\pm$ -- almost until the jet reaches the radius where it attains 
the terminal Lorentz-factor, whereas lower energy photons are free to escape
at the much smaller photospheric radius.

A Poynting jet model for GRBs belongs to this second category, which is
discussed in section 2. The application to Fermi GRBs is presented in \S3. 

\section{High and low energy photon arrival time for a Poynting jet}

The Lorentz factor increases slowly with radius for a magnetic dominated 
jet;  Drenkhahn (2002) finds that $\Gamma\aprop R^{1/3}$
when magnetic field is dissipated via reconnection in a stripped wind 
 and a part of the energy goes into accelerating the jet. The remaining
part of the dissipated magnetic field energy is deposited into particles to
produce a non-thermal distribution which results in a broad-band synchrotron 
spectrum extending to multi-GeV energies. Low energy $\gamma$-rays
can escape the jet when the radiation is produced at a radius that is
larger than the Thomson-photospheric radius. However, high energy $\gamma$-ray
photons can escape only when the jet is at a much
larger radius so that the optical depth for $\gamma$--$\gamma$ 
pair production drops below unity.
Thus, a Poynting jet model for GRBs offers a straightforward explanation
for the delay for the arrival of high energy $\gamma$-ray photons. We 
provide an estimate for the time delay below (\S2.2) after discussing
a few basic results for a Poynting jet (\S2.1).

\subsection{Thomson photosphere and a few basic results for a Poynting jet}

The energy-momentum (EM) tensor for a magnetic outflow is the sum of matter 
and electromagnetic parts. At a distance sufficiently far away from the
central engine so that the magnetic field is in the transverse 
direction\footnote{The transverse component of the magnetic field falls 
off as $r^{-1}$ whereas the radial component as $r^{-2}$. Therefore,
at a large distance from the center of explosion the transverse component
of the magnetic field dominates.} and the thermal
pressure is small, the EM tensor is given by:

\begin{equation}
T^{\mu\nu} = n m_p c^2 u^\mu u^\nu + {1\over 4\pi}
    \left[ \left( u^\mu u^\nu + {1\over 2} g^{\mu\nu} \right)
      B^2 - B^\mu B^\nu \right]  
\end{equation}
where $n$ \& $B$ are number density of protons \& magnetic field strength
in jet comoving frame, $u=\Gamma(1, v,0,0)$ is the 4-velocity, $\Gamma$ is jet 
Lorentz factor,  $B^\mu =$ $^*F^{\mu\nu} u_\nu = (0, 0, B_\theta, B_\phi)$
is the 4-magnetic vector, $^*F^{\mu\nu}=\epsilon^{\mu\nu\alpha\beta}
F_{\alpha\beta}/2$ is the dual-electromagnetic tensor, $F_{\mu\nu} = 
 \partial_\nu A_\mu - \partial_\mu A_\nu$, and $A_\mu$ is 4-potential;
$(B_\theta^2 + B_\phi^2)^{1/2} = B$.

The isotropic luminosity carried by the jet is given by
\begin{equation}
  L = 4\pi r^2 T^{0r} = 4\pi r^2 \left( n m_p c^2 \Gamma^2 v + 
     {B^2 \Gamma^2 v\over 4\pi} \right),
\end{equation}
or
\begin{equation}
  L = 4\pi r^2 n m_p c^2 v \Gamma^2 \left[1 + \sigma(r)\right],   \label{L1}
\end{equation}
where 
\begin{equation}
   \sigma \equiv {B^2 \over 4\pi n m_p c^2},
\end{equation}
is the ratio of magnetic and baryon energy densities. The flux of baryonic 
mass (isotropic equivalent) carried by the jet is
\begin{equation}
    \dot{M} = 4\pi r^2 n m_p \Gamma v
\end{equation}
which should be independent of $r$. Making use of the expression for $\dot{M}$
we can rewrite $L$ as follows
\begin{equation}
   L = \dot{M} c^2 \Gamma (1 + \sigma).   \label{L}
\end{equation}
If radiative losses are small (radiative efficiency for observed $\gamma$-ray
emission is $\lae$50\%) then
\begin{equation}
   L \propto \Gamma (1+\sigma) = {\rm constant}.  \label{La}
\end{equation}

It is easy to see from the equation for $L$ that when the lab frame 
magnetic field strength ($B\Gamma$) decreases with radius faster 
than $r^{-1}$, then the jet Lorentz factor
increases with $r$. For magnetic field undergoing
 reconnection at a fraction of Alfven speed (in jet coming frame)
$\Gamma$ increases as\footnote{We take the Lorentz factor of the jet at 
the base, $R_0$, to be of order unity since it is unphysical for the jet 
to be accelerated instantaneously to a high speed; Drenkhahn (2002) on the
other hand takes $\Gamma(R_0) = \sigma_0^{1/2}$.}
\begin{equation}
\Gamma(r) \sim (r/R_0)^\mu  \quad {\rm for} \quad R_0 < r <R_{sat} \sim R_0 
    \sigma_0^3, \label{gam}
\end{equation}
where $R_0\approx 10^7$cm is the radius where the jet is launched, and
$\mu\approx 1/3$ for the case of reconnection in a striped-magnetic field 
configuration (Drenkhahn, 2002)\footnote{Efficient acceleration of a Poynting
jet can also proceed without reconnection as has been shown in a very nice 
recent work by Granot et al. (2011) who find the same scaling for $\Gamma$ as
in eq. \ref{gam}.}. The jet acceleration might
be more complicated than represented by equation (\ref{gam}), and
the acceleration might cease while $\Gamma\ll\sigma_0$ as suggested by
numerical simulations of highly magnetized jets eg. 
Tchekhovskoy et al. (2008), Komissarov et al. (2010). However, 
the recent work of Granot et al. (2011) also finds $\Gamma\propto R^{1/3}$
after a brief initial spurt of acceleration when $\Gamma$ attains a
value $\sim\sigma_0^{1/3}$. From here on we shall assume that 
equation (\ref{gam}) is valid at least in a limited
radius interval of the Thomson- and pair-production- photospheric radii. 

Radiation can escape the jet when magnetic dissipation occurs at a radius 
larger than the Thomson photospheric radius ($R_p$) which we calculate next.

The optical depth for photon-electron scattering is
\begin{equation}
  \tau_T(R) = \int_R^\infty {dr\over 2\Gamma^2}\,\sigma_T n\Gamma. \label{tau1}
\end{equation}
We assume that there is one electron for every proton; $\sigma_T$ is
Thomson scattering cross-section. The particle
 density $n$ can be expressed in terms of the luminosity ($L$):
\begin{equation}
   n = {\dot{M}\over 4\pi r^2 m_p \Gamma v} \approx
     {L \over 4\pi r^2 m_p \Gamma c^3 \sigma_0}, \label{n}
\end{equation}
where $\sigma_0\equiv\Gamma(R_0)[1 + \sigma(R_0)]$, and we have used equation 
(\ref{L}) to replace $\dot{M}$ in terms of $L$. Substituting equations
(\ref{gam}) \& (\ref{n}) into (\ref{tau1}) we find

\begin{equation}
   \tau_T(R) = {\sigma_T L R_0^{2\mu}\over 8\pi (1+2\mu) m_p c^3 \sigma_0 
    R^{1+2\mu}}.
\end{equation}
Therefore, the photospheric radius, $R_p$, corresponding to $\tau_T(R_p)=1$ is
\begin{equation}
  R_p = \left[ {\sigma_T L R_0^{2\mu} m_p^{-1}c^{-3}\over 8\pi(1+2\mu)\sigma_0}
   \right]^{{1\over 1+2\mu}} = \left[ {6{\rm x}10^{15} L_{52} R_0^{2\mu}\over
       (1+2\mu)\sigma_{0,3} }\right]^{{1\over 1+2\mu}}
   \label{rp}
\end{equation}
where $L_{52} \equiv L/10^{52}$erg s$^{-1}$, $\sigma_{0,3}\equiv\sigma_0/10^3$,
and we use cgs units for numerical values throughout the paper.
For $\mu = 1/3$:
\begin{equation}
\frac{R_P}{R_0} = 1.4\times10^5 L_{52}^{3/5} \sigma_{0,3}^{-3/5} R_{0,7}^{-3/5}
   \label{rpr0}
\end{equation}

The photospheric radius calculated above is modified due to creation of electron-positron pairs by photon-photon collisions. If one were to estimate the 
number density of $e^\pm$ due to this process ($n_\pm$) from the 
observed high-energy gamma-ray spectrum by assuming that all photons 
that have energy larger than $m_e c^2$ in the jet comoving frame are converted 
to pairs as long as these photons are produced at a radius smaller than the 
pair-production photosphere (see \S2.2) one would find that 
$n_\pm$ is larger than $n$ --- the number density of electrons associated
with protons (given by eq. 10) --- by a factor $\sim 50$; under this assumption
$n_\pm/\Gamma\sim n'_\gamma(> 1 MeV*\Gamma)$ which is given by equation (16).
However, this overestimates $n_\pm$ by more than an order of magnitude due
to neglect of an important negative feedback effect. As $e^\pm$ pairs are 
created, the mean thermal Lorentz factor per charged lepton decreases as $\sim 
(1+ n_\pm/n)^{-1}$ if particles are accelerated in shocks or as $\sim 
(1+ n_\pm/n)^{-1/2}$ if particles are accelerated by electric field inside 
the current sheet produced by magnetic-reconnection. The consequence of 
this is that the peak of the spectrum ($E_p$) shifts sharply to lower 
values thereby decreasing the number density of high-energy photons 
capable of pair production (see eq. 16), and that in turn reduces 
further creation of $e^\pm$. Thus, 
pair production is a self-limiting process and it does not allow 
charge-lepton density to be increased by 
more than a factor a few, and so the Thomson-photosphere radius given by 
equation (12) is not in error by more than a factor $\sim 5$; this 
error in $R_p$ contributes $\lae20$\% error in the estimate of arrival time
delay for high-energy $\gamma$-ray photons (see eq. 22)\footnote{If pair
production were to change particle density by a large factor then we should
see a big jump in $E_p$ when the jet crosses the pair-production photosphere.
Fermi/GBM sees a small increase to $E_p$ a few seconds after the trigger, and 
that suggests that pair loading has a small effect on electron density.}.

A consequence of the pair production process (described above) is that the
spectral-peak shifts to lower energies and the high-energy spectrum softens. 
Therefore, when the jet crosses 
the radius where the pair-production opacity drops below unity, high-energy
$\gamma$-rays ($>$10$^2$ MeV) are able to escape conversion to $e^\pm$,
and the spectral peak shifts to higher energies and the high-energy spectrum
hardens. Fermi/LAT data shows that the arrival of high-energy photons is
accompanied by these spectral changes, eg. Abdo et al. 2009a.

\subsection{Pair-production photosphere for a Poynting jet and the delay 
   for the arrival of GeV photons}

Consider a photon of energy $E_0$ in observer frame. Its energy in the
jet comoving frame ($E_0'$), and the minimum photon energy ($E'_\pm$)
needed to convert this photon to an electron-positron pair are given by
\begin{equation}
  E_0' = (1+z) E_0/\Gamma, \quad E'_\pm \approx m_e^2 c^4/E_0',
  \label{Epm}
\end{equation}
where $z$ is the GRB redshift, and $m_e$ is electron mass. The
comoving number density of photons of energy $>E'_\pm$ --- 
$n'_\gamma(>E'_\pm)$ --- can be calculated from the observed $\gamma$-ray 
luminosity, $L_\gamma(E)$.
Let us consider the observed $\gamma$-ray spectrum to peak at energy
$E_p$. The spectrum above $E_p$ is a powerlaw function with photon index 
$\beta$, and the frequency-integrated-luminosity above $E_p$ is $L_{>p}$,
i.e. $L_{>p} \equiv \int_{E_p}^\infty dE\, L_\gamma(E)\propto E_p^{2-\beta}$. 
The comoving number density of photons with observer frame energy 
$\ge E>E_p$, at radius $R$, is given by

\begin{equation}
   n'_\gamma(>E) = {1\over 4\pi R^2\Gamma} \int_E^\infty dE \,
     {L_\gamma(E) \over E(1+z)}. 
\end{equation}
Or
\begin{equation}
   n'_\gamma(>E) = {1\over 4\pi R^2\Gamma} \left( {\beta - 2\over \beta - 1} 
   \right) \left[ {E_p\over E}\right]^{\beta-1} { L_{>p} \over (1+z)E_p c}.
     \label{nphoto}
\end{equation}

The optical depth for a photon of energy $E_0$ to get converted to $e^\pm$
while traversing through the jet starting from a radius $R$ is given by 

\begin{equation}
   \tau_\pm(E_0, R) \approx \sigma_{\gamma\gamma} \, n'_\gamma(>E_\pm)\,
  [R/\Gamma],
\end{equation}
where $\sigma_{\gamma\gamma} = 6\times10^{-26}$ cm$^2$ is the 
photo-pair-production cross-section just above the photon threshold energy 
for producing $e^\pm$,
and $R/\Gamma$ is the comoving radial width of a causally connected region.
Using equation (\ref{nphoto}) for comoving photon density, and equation
(\ref{Epm}) for $E_\pm=\Gamma E'_\pm/(1+z)$ we find

\begin{equation}
   \tau_\pm \approx  \left( {\beta - 2\over \beta - 1} \right)
     {\sigma_{\gamma\gamma}\over 4\pi R\Gamma^2} {L_{>p} \over (1+z)^{3-2\beta}
    E_p c} \left[ { E_p E_0 \over \Gamma^2 m_e^2 c^4} \right]^{\beta -1}.
\end{equation}

Substituting for $\Gamma$ from equation (\ref{gam}) we find the
radius $R_{\gamma\gamma}(E_0)$ where $\tau_\pm$ drops below unity 
so that photons of energy $E_0$ are able to escape conversion to pairs --

\begin{equation}
\frac{R_{\gamma\gamma}(E_0)}{R_0} = \Bigg[ \frac{\beta-2}{\beta-1}
   \frac{L_{>p}\sigma_{\gamma\gamma} E_p^{\beta-2} E_0^{(\beta-1)} R_0^{-1} }
       {4\pi c(m_e c^2)^{2(\beta-1)} (1+z)^{3-2\beta}}\Bigg]^{\frac{1}
       {1+2\mu\beta}}   \label{rgamma}
\end{equation}
For $\mu$ = 1/3 and $\beta$=2.2:
\begin{equation}
\frac{R_{\gamma\gamma}(E_0)}{R_0} \simeq 4.1{\rm x}10^6 L_{>p,52}^{0.41}
    E_{p,-6}^{0.08} E_{0,-4}^{0.49} R_{0,7}^{-0.41}(1+z)^{0.57}, \label{rgamma1}
\end{equation}
where $E_{p,-6}$ is photon energy at the peak of the observed 
spectrum in units of 1 MeV \& $E_{0,-4}$ is the high energy $\gamma$-ray
photon of energy in unit of 100 MeV for which the escape radius 
($R_{\gamma\gamma}$) is calculated.

Photons of energy less than about $(R_p/R_0)^{\mu}/(1+z)$ MeV $\sim20$MeV
are not much affected by pair conversion considerations and we observe these
photons essentially unattenuated whenever they are 
generated at a radius larger than the photospheric radius; for this estimate
we used equation (\ref{rpr0}) for the photospheric radius $R_p$, took 
$\mu=1/3$, and assumed that the spectral peak $E_p\sim 1$ MeV.
However, photons of energy $\ge$10$^2$MeV are unable to escape until the 
jet has propagated to a much larger radius of $R_{\gamma\gamma}$.

The time it takes for the jet to propagate from $R_p$ to 
$R_{\gamma\gamma}(E_0)$, as measured by the arrival of photons at an 
observer from these radii, is the observed delay for photons of 
energy $E_0$. This delay in observer frame, $\Delta t(E_0)$, is 
straightforward to calculate and is given by

\begin{equation}
\Delta t(E_0) = (1+z) \int_{R_p}^{R_{\gamma\gamma}(E_0)} {dr \over 
   2c \Gamma^2}.
\end{equation}
Using equation (\ref{gam}) for $\Gamma$, the above integral reduces to
\begin{equation}
\Delta t \simeq \frac{R_0(1+z)}{2c(1-2\mu)}\Bigg[\Bigg(
   \frac{R_{\gamma\gamma}(E_0)}{R_0}\Bigg)^{1-2\mu}-\Bigg(
   \frac{R_p}{R_0}\Bigg)^{1-2\mu}\Bigg]
   \label{deltat}
\end{equation}

We use equation (\ref{deltat}) together with equations (\ref{rp}) \& 
(\ref{rgamma}) to calculate the expected delay for the arrival of 
$>$10$^2$MeV photons (in comparison to photons of energy $\lae10$ MeV).

It should be noted that according to the reconnection model considered here
the observed $\gamma$-ray luminosity is roughly proportional to the rate 
of dissipation of magnetic energy. The jet luminosity carried by magnetic fields is
\begin{equation}
  L_B = B^2\Gamma^2 R^2 v = L - \dot{M}\Gamma c^2 = \sigma\Gamma \dot{M} c^2
   \approx L \left( 1 - \sigma^{-1} \right),
\end{equation}
where $L$ is the rate of energy transport by the jet, and $B$ is magnetic 
field in jet comoving frame. The $\gamma$-ray luminosity ($L_\gamma$),
when the jet is above the photosphere, can be calculated by estimating the 
total magnetic energy dissipated ($\Delta E_B$) between radii $R$ \& 2$R$ 
during time interval $\delta t\sim R/(2c\Gamma^2)$ -- in observer frame -- when 
the jet radius roughly doubles
\begin{equation}
   \Delta E_B \sim R{d L_B\over dR} \delta t \sim {\mu L \Gamma \delta t\over 
   \sigma_0}, \quad L_\gamma \sim {\Delta E_B\over \delta t} 
   \propto t^{{\mu\over 1 - 2\mu}},
\end{equation}
where $t$ is time in observer frame. For $\mu=0.3$, $\langle 
L_\gamma\rangle\propto t^{3/4}$ while the jet propagates between the 
photosphere and the radius where $\Gamma(R)\sim\sigma_0$, i.e. for $1s 
\lae t\lae 4$s\footnote{$\langle L_\gamma\rangle$ is the average of 
$L_\gamma$ which is subject to possibly large
fluctuations due to central engine activity, stochastic magnetic reconnection 
and relativistic outflow produced in the layer where magnetic dissipation takes place.}; The temporal behavior of $L_\gamma$ on a longer
time scale is governed by the activity of the central engine.

\section{Application to Fermi bursts}
\medskip

\begin{table*}
%\begin{center}
\begin{minipage}{\textwidth}
\begin{center}
\begin{tabular}{ l r c c c c c c }
\\
\hline\hline

GRB \# &  E$_p$ & E$_{\mathrm{iso,54}}$  &  T$_{90}$ &  z & $\Delta$t$_{obs}$ & $\Delta$t$_{th}$ ($\mu=1/3$) & $\Delta$t$_{th}$ ($\mu=0.3$) \\
   &  keV & erg & s & &  s  &  ~s ~~~($R_0$=3x10$^7$) & s ($R_0$=1.5x10$^7$) \\ [0.5ex]

\hline

080916C$^1$  & 424 & 8.8  & 66 & 4.35 & 4 & 2.3 & 4.7 \\
090323$^2$ & 812  & $>$3 & 150 & 3.57 & $\sim$5 & $>$1.2 & $>$3 \\
090510$^3$ & 3900   & 0.11 & 0.6 & 0.90 & 0.15 & 0.5 & 1.4 \\
090902B$^4$ & 726 & 3.7 & 22 &  1.82 & 2.5 & 0.8 & 2.2  \\
090926A$^5$ & 259 & 2.2 & 13 & 2.11 & 3 & 0.9 & 2.4  \\
\hline
\end{tabular}
\end{center}
\end{minipage}
\caption{{\small GRBs detected by Fermi/LAT with known redshifts for which
 $>$10$^2$MeV photons are observed to arrive after lower energy $\gamma$-rays.  080916C, 090323, 090902B \& 090926A are long-GRBs whereas 090510 is a short-GRB. $E_p$ is the photon 
energy at the peak of $\nu f_\nu$ spectrum, $E_{iso,54}$ is the isotropic 
equivalent of total energy radiated in $\gamma$-rays in unit of 10$^{54}$ ergs,
$\Delta$t$_{th}$ is the theoretically calculated delay (using eq. \ref{deltat})
for the arrival of 10$^2$ MeV photons, and $\Delta$t$_{obs}$ is the observed
delay. We used $\beta=2.2$, $\sigma_0=10^3$ \& ($\mu, R_0$) = ($1/3, 3\times10^7$cm) or ($0.3, 1.5\times10^7$cm) for $\Delta$t$_{th}$ calculations.  
{\bf References:} (1) GRB 080916C: E$_p$,  T$_{90}$ - van der Horst \& Goldstein (2008), E$_{iso}$ - Abdo et al. (2009a),  z - Greiner et al. (2009); (2) GRB 090323 -- Zhang et al. (2010); (3) GRB 090510: E$_p$, E$_{iso}$, T$_{90}$, $\Delta$t$_{\mathrm{obs}}$ - Ackermann et al. (2010), z - McBreen et al. 2010; (4) GRB 090902B: E$_p$, E$_{iso}$, T$_{90}$ - Abdo et al. (2009b), z - Cucchiara et al. (2009); (5) GRB 090926A: E$_p$, E$_{iso}$, T$_{90}$, $\Delta$t$_{\mathrm{obs}}$ - Ackermann et al.
(2011) \& Zhang et al. (2011), z - Malesani et al. (2009).  }}
\end{table*}

Fermi has detected 17 bursts with photons of energy $>$10$^2$MeV. Five
of these bursts have redshift measurements and good photon statistics 
in the LAT band to determine accurately the delay for the arrival of 
$>$10$^2$MeV photons with respect to the Fermi/GBM trigger time\footnote{Two bursts -- GRB 091003 \& 100414A -- were detected by LAT and have known redshift but these bursts do not show any measurable delays in the arrival of $>$10$^2$MeV photons and these are not considered in this work; the jet for these bursts, perhaps, might not be magnetic dominated or $R_p\sim R_{\gamma\gamma}$.}. 
For these bursts we carry out a comparison between the expected and observed delays.

A few basic properties of these five bursts are presented in Table 1.
The table also contains the observed and the expected delays for each
of these bursts. All of the expected delays, reported in the column marked
$\Delta t_{th}$, were calculated for 100 MeV photons using the observed 
spectral peak ($E_p$) and luminosity ($E_{iso}/T_{90}$) for each burst.

The theoretically calculated delay for the four long-GRBs for $\mu=1/3$ is 
smaller than the observed delays by a factor $\sim3$ (see Table 1). 
The expected delay has a very weak
dependence on $\gamma$-ray luminosity ($L_{>p}^{0.14}$), $E_p$ \& $\beta$
(see equations \ref{rgamma1} \& \ref{deltat}). The delay depends primarily on
$\mu$, $z$ and $R_0$; $\Delta t_{th}$ increases almost linearly with $R_0$
\& $z$ and the dependence on $\mu$ is very strong.

A factor of a few difference between $\Delta t_{th}$ and $\Delta t_{obs}$ could
 be due to the fact that $R_0$ is a little larger than the value we have assumed for
the calculations reported in table 1 for $\mu=1/3$. If $R_0$ were to be larger by a factor
of $\sim 2$ for long-GRBs and smaller by a factor $\sim 3$ for short-GRBs
then $\Delta t_{th} \approx \Delta t_{obs}$. The radius at which jet is
launched ($R_0$) depends on the mass of the central blackhole produced
in these explosions, and it is not surprising that long- and short- GRBs
leave behind blackholes of different mass.

The value of $\mu=1/3$ is motivated by the analytical 
results of Drenkhahn (2002) for jet acceleration via magnetic reconnection 
for an alternating-field configuration. The numerical results presented
by Drenkhahn \& Spruit (2002) --- see fig. 1 --- show that the average $\mu$
might be a little bit smaller than $1/3$. Moreover, the value of $\mu$ 
depends on the magnetic field geometry --- whether the magnetic field 
gradient is larger in the radial or transverse direction (Drenkhahn 2002, fig. 3) ---
and therefore there is some uncertainty in regards to the precise value 
$\mu$ might take for a Poynting jet. If we were to take $\mu=0.3$ instead 
of $1/3$ (and $R_0=1.5\times10^7$cm) then the expected delays for long-GRBs
are approximately equal to their observed values (see Table 1 -- column
marked $\Delta t_{th}, \mu=0.3$). However, for the short GRB (090510) the 
discrepancy is a factor 10 for $\mu=0.3$ \& $R_0=1.5\times10^7$cm (table 1); 
this discrepancy disappears if we take $R_0\sim10^6$cm which might be 
reasonable for a short burst produced by a binary neutron star merger that 
gives birth to a blackhole of 2--3 $M_\odot$.

For $\sigma_0=10^3$ the radius where the magnetization parameter drops
below unity, i.e. the magnetic dissipation becomes insignificant and jet 
acceleration ceases, is $R_{sat} \sim R_0 \sigma_0^3 \sim 10^{16}$cm. 
The pair-production photosphere radius ($R_{\gamma\gamma}$) for the 
long-GRBs considered in Table 1 is $\sim 10^{15}$cm. Thus, 
$R_{\gamma\gamma}<R_{sat}$ and the calculations presented in
\S2 are applicable to the GRBs considered in this paper. 

We note that according to the Poynting jet model analyzed here the delay 
for the arrival of high-energy photons increases with increasing photon 
energy as $\sim E_0^{0.17}$. This can be used as a test of this model.

\section{Discussion}
\medskip

The external forward shock model for GRBs is in good agreement with the
observed Fermi data for the high energy photons (energy $\gae10^2$MeV) 
after the prompt phase ($t\gae30$s) as well as the x-ray, optical and radio data (Kumar \& Barniol Duran, 2010). 
However, the external-forward-shock model cannot account for the
high-energy data during the prompt phase if the fluctuations in the
lightcurve are on a short time-scale (less than $\sim$1s ) and 
are correlated with $<10$ MeV lightcurve. In this case one needs to 
look for another mechanism for generation of high-energy photons 
during the prompt phase that can explain the observed delay of
a few seconds reported by Fermi for a number of GRBs (Abdo et al. 2009a;
Abdo et al. 2009b). We note that it has been known for a long time 
that there is a slight time difference in the arrival of low and 
high-energy $\gamma$-ray photons of energy less than $\sim 10$ MeV.
However, when photons of energy less than $\sim 10$ MeV are considered, it is 
found that lower energy $\gamma$-ray photons lag higher energy photons (Norris 
et al. 1996), which is opposite to the result reported by Fermi for 
$>$10$^2$MeV photons. Thus, the observed delay in the arrival of $>$10$^2$MeV 
photons must have a distinct origin than that for lower energy photons.

A number of mechanisms have been suggested for this delay, eg. Razzaque et al.  (2010), Asano et al. (2009), Toma et al. (2009), Vurm et al. (2011), Bo\v snjak et al. (2009), Daigne et al. (2011), M\'esz\'aros and Rees (2011).

We report in this work that the dissipation of magnetic fields in a Poynting 
jet model for GRBs offers a natural explanation for the observed delay in 
the time of arrival of photons of energy $>$10$^2$MeV. This delay 
arises because the Lorentz factor of a Poynting jet increases
slowly with radius ($\Gamma\aprop R^{1/3}$) and as a result
high energy photons are converted to $e^\pm$ pairs even when
they are produced far above the Thomson-photosphere radius ($R_p$) 
whereas lower energy photons ($E\lae10$ MeV) can escape readily starting at 
$R_p$. A straightforward calculation shows (\S2) that the delay for the arrival
of $>$10$^2$MeV photons due to this process is similar to the observed value.

The recent work of M\'esz\'aros and Rees (2011) also considered a Poynting jet 
(with $\Gamma$ varying laterally) for
explaining the delay. However, according to them the diffusion of neutrons 
from the outer part of the jet to the faster-moving, inner part,
and the resulting collisions between protons and neutrons were responsible 
for the generation of delayed GeV photons. 

What we have shown here is that even without a neutron component to a 
Poynting jet the high-energy-photon delay can be understood ---
the acceleration and generation-of-radiation for a highly magnetized jet are
coupled processes, and this offers a simple explanation for the observed delay.

The Poynting jet model described here predicts that the delay should
depend primarily on $R_0$, $z$ (linearly) \& $\mu$, and it has a weak
dependence on high-energy photon energy ($E_0^{0.17}$), $E_p$ ($\aprop 
E_p^{0.05}$) \& $L_{>p}^{0.14}$. 
If the average $\mu$ does not vary from one burst to another 
(which is determined by the magnetic field topology of the jet) then one 
can use the delay in arrival of $>$10$^2$MeV photons to determine 
the product of burst redshift and $R_0$.

\section*{Acknowledgments}

This work was carried out while the authors were visiting Universit\`a di 
Ferrara. We are grateful to Professor Filippo Frontera for his kind
hospitality. PK thanks Rodolfo Barniol Duran for helpful discussions.
This work has been funded in part by NSF grant ast-0909110,
and a Fermi-GI grant. ZB acknowledges the French Space Agency (CNES) for financial support.

\end{document}